\documentclass[10pt,a4paper,oneside]{article}
\usepackage[a4paper, margin=20mm, total={210mm, 297mm}]{geometry}
\usepackage[margin=0pt,labelfont={bf}]{caption}
\DeclareCaptionLabelSeparator{dot}{. }
\usepackage{apacite}    
\usepackage[utf8]{inputenc} 

\usepackage{fancyhdr}
\def\lhead{}
\def\rhead{}
\fancypagestyle{firstpage}
{
	\fancyhead[L]{\lhead}    
	\fancyhead[R]{\rhead}
	\fancyfoot[L]{}
}
\newcommand*{\citet}[1]{\citeauthor{#1} \citeyear{#1}}
\usepackage{mathtools,amssymb}
\usepackage{dsfont}
\usepackage{color}
\usepackage[dvipsnames]{xcolor}
\usepackage{caption}
\usepackage{subcaption}
\usepackage{float}
\usepackage{color}
\usepackage{tikz}
\usepackage{pgfplots}
\usepackage[inline]{enumitem}
\definecolor{ao(english)}{rgb}{1, 0.1, 0.1}

\newcount\liczpkt
\long\def\pkt#1{\global\advance\liczpkt by1%
	\begingroup
	\parindent=0pt
	{\bf \the\liczpkt.} #1 \medbreak
	\endgroup}

\def\Y{{\it YES\ }}

\def\Nd{{\it NO}}
\def\Jdn{\mathds{1}}

\def\piCM{\hat\pi_{CM}} 
\def\piC{\tilde\pi_{C}} 
\def\c{{C(\pi)}}

\def\LeftCi{\left(B^{-1}\left(\lceil u\rceil+1,n-\lceil u\rceil;\frac{1+\delta}{2}\right)-(1-q)\right)/(2q-1)}
\def\RightCi{\left(B^{-1}\left(\lfloor u\rfloor,n-\lfloor u\rfloor+1;\frac{1-\delta}{2}\right)-(1-q)\right)/(2q-1)}
\def\DeltaCi{
	\dfrac{B^{-1}\left(\lfloor u\rfloor,n-\lfloor u\rfloor+1;\frac{1-\delta}{2}\right)-B^{-1}\left(\lceil u\rceil+1,n-\lceil u\rceil;\frac{1+\delta}{2}\right)}{2q-1}}

\def\condL{B(\lceil u\rceil+1,n-\lceil u\rceil;1-q)<\frac{1+\delta}{2}}
\def\condR{B(\lfloor u\rfloor,n-\lfloor u\rfloor+1;q)>\frac{1-\delta}{2}}
\pgfplotsset{
		myfigureformat100/.append style={
		 tick label style={font=\footnotesize}, 
		axis y line=left,
		axis x line=bottom,
		ymin=0.92,ymax=0.985,
	height=4.5cm,
	width=7.5cm,
	}}
\pgfplotsset{
	myfigureformat1000/.append style={
		yticklabel style={/pgf/number format/.cd,fixed,fixed zerofill,precision=3,}, 
		tick label style={font=\footnotesize}, 
		axis y line=left,
		axis x line=bottom,
		ymin=0.94,ymax=0.96,
		height=4.5cm,
		width=7.5cm,
		try min ticks=5,
}}
\pgfplotsset{
	myfigureformat-pivsq/.append style={
		yticklabel style={/pgf/number format/.cd,fixed,fixed zerofill,precision=1,
		}, 
		tick label style={font=\footnotesize}, 
		height=4.5cm,
		axis y line=left,
		axis x line=bottom,
		ymin=0,ymax=1,
}}

\newenvironment{penum}{
	\begin{enumerate}[label={\alph*})]
		\setlength{\topsep}{1pt}
		\setlength{\itemsep}{1pt}
		\setlength{\parskip}{0pt}
		\setlength{\parsep}{0pt}
	}{\end{enumerate}}	
\newenvironment{penum*}{
	\begin{enumerate*}[label={\alph*})]
	}{\end{enumerate*}}
\newenvironment{enum}{
	\begin{enumerate}[label={\arabic*)}]
		\setlength{\topsep}{1pt}
		\setlength{\itemsep}{1pt}
		\setlength{\parskip}{0pt}
		\setlength{\parsep}{0pt}
	}{\end{enumerate}}	
\newbox\tabstrutbox
\setbox\tabstrutbox=\hbox{\vrule height12pt depth4pt width0pt}
\def\tb{\relax\ifmmode\copy\tabstrutbox\else\unhcopy\tabstrutbox\fi}

\font\kapitaliki=plcsc10 

\title{The Optimal Sample Size\\ in Crosswise Model for Sensitive Questions\\}
\date{}

\author{
	Stanisław Jaworski\footnote{{\kapitaliki Corresponding author}: Stanisław Jaworski, Department of Econometrics and Statistics, Warsaw University of Life Sciences, Nowoursynowska 159, PL-02-767 Warsaw}\\
	Wojciech Zieli\'{n}ski\\
	Warsaw University of Life Sciences (Poland)\\
	e-mail: stanislaw$\_$jaworski@sggw.edu.pl\\		
	e-mail: wojciech$\_$zielinski@sggw.edu.pl\\	}	

\begin{document}
	\maketitle
	\begin{abstract}
		{\noindent The problem is in the estimation of the fraction of population with a stigmatizing characteristic. In the paper the nonrandomized response model proposed by Tian, Yu, Tang, and Geng (2007) is considered. The exact confidence interval for this fraction is constructed. Also the optimal sample size for obtaining the confidence interval of a given length is derived. \\
		}
	\end{abstract}
	{\kapitaliki Keywords}:  sensitive questions, NNR model, exact confidence interval\\
	{\kapitaliki 2010 Mathematics Subject Classification}:  62F25, 62P20\\

\section{Introduction}

The problem is in the estimation of the  percentage of population who ``committed''  socially stigmatizing ``crimes'' such as corruption, tax frauds, illegal work (black market), drug uses, violence against children and  other.  

Mathematicaly, let $Y$ be a random variable such that $$P\{Y=1\}=\pi=1-P\{Y=0\}.$$ The r.v. takes on the value $1$ when the answer to the sensitive question is \Y and the value $0$ otherwise. The number $\pi\in(0,1)$ is the probability of the positive answer to the sensitive question, i.e. $\pi\cdot100\%$ is the percentage of interest. We want to estimate the probability $\pi$, i.e. we are going to construct a confidence interval for $\pi$.

Let $Y_1,\ldots,Y_n$ be a sample. The statistical model for the sample is
		$$\left(\{0,1,\ldots,n\},\{Bin(n,\pi),\pi\in(0,1)\}\right),$$
where $Bin(\cdot,\cdot)$ denotes a Binomial distribution.

The difficulty which arises is such that random variables $Y_1,\ldots,Y_n$ are not observable. Answers to the sensitive question are ``hidden'' through asking a ``neutral'' question, which is answered \Y or \Nd. It is assumed that the ``neutral'' question is independent from the sensitive question. In a questionnaire two questions are asked: sensitive and neutral. But only one answer is registered and the interviewer does not know which of the two questions the interviewee answered.

The first method of obscuring the answer to a sensitive question was proposed by (Warner, 1965). His method consists in the randomization of answers. This randomization is done by the respondent and the interviewer does not know what the answer to the sensitive question is. This model was extended in different ways (Horvitz,
Shah, \& Simmons, 1967; Greenberg, Abul-Ela, \& Horvitz, 1969; Raghavarao, 1978; Franklin, 1989; Arnab,
Shangodoyin, \& Arcos, 2019; Arnab, 1990, 1996; Kuk, 1990; Rueda, Cobo, \& Arcos, 2015).

(Tian et al., 2007) proposed a nonrandomized response model (NRR). Their idea consists in asking two questions simultaneously: one sensitive and one neutral. This model was extended to other, similar approaches: (Yu,
Tian, \& Tang, 2008; Tan, Tian, \& Tang, 2009; Tian, 2014).

Unfortunately, the problem of constructing confidence intervals for $\pi$ was considered rather rarely. Moreover, proposed confidence intervals are asymptotic. These confidence intervals are not c.i. in (Neyman, 1934, p. 562) sense: they do not keep prescribed confidence level. In what follows the finite sample size confidence interval is proposed. Its construction is based on the distribution of the Maximum Likelihood estimator of $\pi$. We consider only the crosswise model proposed by (Yu et al., 2008).

In section 2. we give the method of the construction of a new confidence interval for $\pi$. In section 3. we recall the construction of asymptotic confidence intervals. We also discuss the probability of the coverage of presented confidence intervals. In the next section we present the methods of sample size selection. This section plays the main role in our paper. In section 5. some concluding remarks are given.

\section{Confidence interval in Crosswise Model} 

In the crosswise model ($CM$)  respondents are presented with two questions simultaneously, one neutral and one sensitive. They are instructed to report $1$ only if answers to both questions are the same, i.e. the observable variable in this model is $Z$, where 
$$Z=\begin{cases}
	1,&\hbox{if both answers are \Y or \Nd}, \\
	0,&\hbox{otherwise.} \\
\end{cases}$$
The answers of $n$ respondents may be treated as the realizations of a Binomial distribution with parameters $(n,\varrho)$, where $\varrho$ is the probability of receiving an outcome 1 of the $Z$ variable. Assume that the asked questions are independent and the probability of the answer $\Y$ to the sensitive question is $\pi$ ($q$ for the neutral question). It is assumed that $q$ is known.
Hence, in $CM$ model $$\varrho=q\pi+(1-q)(1-\pi)=(2q-1)\pi+(1-q).$$ In this model $$\pi=\frac{\varrho-(1-q)}{2q-1}.$$ 
Without loss of generality we  assume that $q<0.5$. 

Let $Z_1,\ldots,Z_n$ be a sample. MLE of $\varrho$ is $\hat\varrho=\frac{1}{n}\sum_{i=1}^nZ_i$. The distribution of $n\hat\varrho$ is $Bin(n;\varrho)$.

The MLE of $\pi$ has the form
$$\piCM=\max\left\{\min\left\{\frac{\hat\varrho-(1-q)}{2q-1},1\right\},0\right\}.$$
Let $Bin\left(\cdot,n;\varrho\right)$ denote the CDF of the binomial distribution with the probability of success equal to $\varrho$ and let $B(a,b;\cdot)$ denote the CDF of the Beta distribution with parameters $(a,b)$. 

In the derivation of the pdf of $\piCM$ the following known relationship will be applied: if $\xi$ is a random variable distributed as binomial with parameters $(n,\rho)$ then
$$P_\rho\{\xi\leq x\}=\sum_{i=0}^n\binom{n}{i}\rho^i(1-\rho)^{n-i}=B(n-x,x+1;1-\rho).$$

The pdf of the distribution of $\piCM$ is
$$\begin{aligned}
	P_\pi\left\{\piCM= x\right\}
	&=\begin{cases}
		P_\pi\left\{n\hat\varrho\geq\lceil(1-q)n\rceil\right\},&\hbox{for $x=0$},\\
		\binom{n}{\lceil u\rceil}((2q-1)\pi+(1-q))^{\lceil u\rceil}(1-(2q-1)\pi-(1-q))^{n-\lceil u\rceil},&\hbox{for $0< x<1$}\\
		P_\pi\left\{n\hat\varrho\leq\lfloor qn\rfloor\right\},&\hbox{for $x=1$}\\
	\end{cases}\\
	&=\begin{cases}
		B\left(\lceil n(1-q)\rceil,n-\lceil n(1-q)\rceil+1;(2q-1)\pi+(1-q)\right),&\hbox{for $x=0$}\\
		\binom{n}{\lceil u\rceil}((2q-1)\pi+(1-q))^{\lceil u\rceil}(1-(2q-1)\pi-(1-q))^{n-\lceil u\rceil},&\hbox{for $0< x<1$}\\
		1-B\left(\lfloor nq\rfloor+1,n-\lfloor nq\rfloor;(2q-1)\pi+(1-q)\right)&\hbox{for $x=1$}\\
	\end{cases}\\
\end{aligned},$$
where $u=n(x(2q-1)+(1-q))$. Here $\lceil x\rceil$ denotes the smallest integer not smaller than $x$ and $\lfloor x\rfloor$ denotes the greatest integer not greater than $x$.

The CDF of $\piCM$ has the form:
$$F_\pi(x)=P_\pi\left\{\piCM\leq x\right\}=\begin{cases}
		B(\lceil u\rceil,n-\lceil u\rceil+1;(2q-1)\pi+(1-q)),&\hbox{for $0\leq x<1$},\\
		1,&\hbox{for $x=1$}.\\
	\end{cases}\\
$$
Since the distribution of $\piCM$ is  discrete we have
$$P_\pi\left\{\piCM<x\right\}=\begin{cases}
	0,&\hbox{for $x=0$},\\
	B(\lfloor u\rfloor +1,n-\lfloor u\rfloor ;(2q-1)\pi+(1-q)),&\hbox{for $0< x\leq1$}.\\
\end{cases}\\
$$

Note that the family $\{F_\pi,\pi\in[0,1]\}$ is stochastically ordered, i.e. 
$$F_{\pi_1}(\cdot)\geq F_{\pi_2}(\cdot),\quad\hbox{for\ }\pi_1<\pi_2.$$

Let $\delta$ be a given confidence level and let $\piCM=x$ be the observed value of $Z$. The equitailed confidence interval $\left(\pi_L(x;\delta);\pi_R(x;\delta)\right)$ for $\pi$ is defined as
$$\begin{cases}
    \pi_L(x;\delta)=\arg\inf_{\pi}P_\pi\left\{\piCM\leq x\right\}\geq \frac{1+\delta}{2},\\
    \pi_R(x;\delta)=\arg\sup_{\pi}P_\pi\left\{\piCM< x\right\}\leq \frac{1-\delta}{2}.\\
\end{cases}$$
The function $\pi\to F_\pi(x)$ for a given $x$ has two jumps. The first one is at $\pi=0$ and the second one is at $\pi=1$. At $\pi=0$ there is a jump equal to $1-B(u,n-u+1;1-q)$, and at $\pi=1$ the jump equals $B\left(u,n-u+1;q\right)$. Hence the confidence interval for $\pi$ has the form
$$\begin{aligned}
\pi_L(x;\delta)&=
\begin{cases}
	0,&\hbox{if $\condL$},\\
	\LeftCi,&\hbox{otherwise},\\
\end{cases}\\
\pi_R(x;\delta)&=
\begin{cases}
	1,&\hbox{if $\condR$},\\
\RightCi,&\hbox{otherwise}.\\
\end{cases}
\end{aligned}
\eqno{(CP)}$$

The coverage probability of the above confidence interval is, by the definition, greater or equal to the given confidence level. In Figures 1 and 2 the coverage probability is presented for $n=100$ and $n=1000$, respectively (solid line). The confidence level is assumed to be $0.95$. This coverage probability is calculated, not simulated.

\section{Asymptotic c.i.}

The most common method of constructing c.i. is the application of Central Limit Theorem. This method was applied by (Yu et al., 2008) and (Tian, 2014).

If $\xi$ is a random variable distributed as Binomial with parameters $n$ and $\varrho$, then, by CLT the distribution of $\xi$ tends in distribution as $n$ tends to infinity to normal variate with the mean value $n\varrho$ and the variance $n\varrho(1-\varrho)$. So 
$$\hat\varrho\sim AN\left(\varrho,\frac{\varrho(1-\varrho)}{n}\right).$$

(Tian, 2014) considered the following estimator of $\pi$:
$$\piC=\frac{\hat\varrho-(1-q)}{2q-1}.$$
Its properties are as follows:
$$\begin{aligned}
	&E_\pi\piC=\pi,\\
	&Var_\pi(\piC)=\frac{\varrho(1-\varrho)}{n(2q-1)^2}=\frac{\pi(1-\pi)}{n}+\frac{q(1-q)}{n(2q-1)^2}.\\
\end{aligned}$$

The unbiased estimator of the variance of $\piC$ is given by the formula:
$$\widehat{Var\piC}=\frac{\hat\varrho(1-\hat\varrho)}{(n-1)(2q-1)^2}=\frac{\piC(1-\piC)}{n-1}+\frac{q(1-q)}{(n-1)(2q-1)^2}.$$

By CLT we have
$$\piC\sim AN\left(\pi,\frac{\pi(1-\pi)}{n}+\frac{(1-q) q}{n(2q-1)^2}\right).$$
Hence
$$\frac{\piC-\pi}{\sqrt{Var\piC}}\sim N(0,1).$$
There are two ways of constructing c.i. on the basis of the above approximation. 

The first one is based on the solution of the inequality
$$\left|\frac{\piC-\pi}{\sqrt{Var\piC}}\right|<u_{\frac{1+\delta}{2}}$$
or equivalently
$$\left(\frac{\piC-\pi}{\sqrt{Var\piC}}\right)^2<u^2_{\frac{1+\delta}{2}}.$$
Solving the inequality with respect to $\pi$ we obtain the following c.i.
$$\frac{2 n\piC + u_{\frac{1+\delta}{2}}^2 \pm u_{\frac{1+\delta}{2}}\sqrt{4n \piC(1-\piC) + \frac{4 n (1 - q) q + u_{\frac{1+\delta}{2}}^2}{(1 - 2 q)^2}}}{2\left(n +u_{\frac{1+\delta}{2}}^2\right)}.\eqno{(WP)}$$
In the second approach the variance of $\piC$ in CLT is substituted by its unbiased estimator
$$\frac{\piC-\pi}{\sqrt{\widehat{Var\piC}}}\sim N(0,1).$$
Solving the inequality with respect to $\pi$
$$\left|\frac{\piC-\pi}{\sqrt{Var\piC}}\right|<u_{\frac{1+\delta}{2}}$$
the following c.i. is obtained
$$\left(\piC-u_{\frac{1+\delta}{2}}\sqrt{\widehat{Var\piC}};\ \piC+u_{\frac{1+\delta}{2}}\sqrt{\widehat{Var\piC}}\right).\eqno{(AP)}$$

\begin{figure}[H]
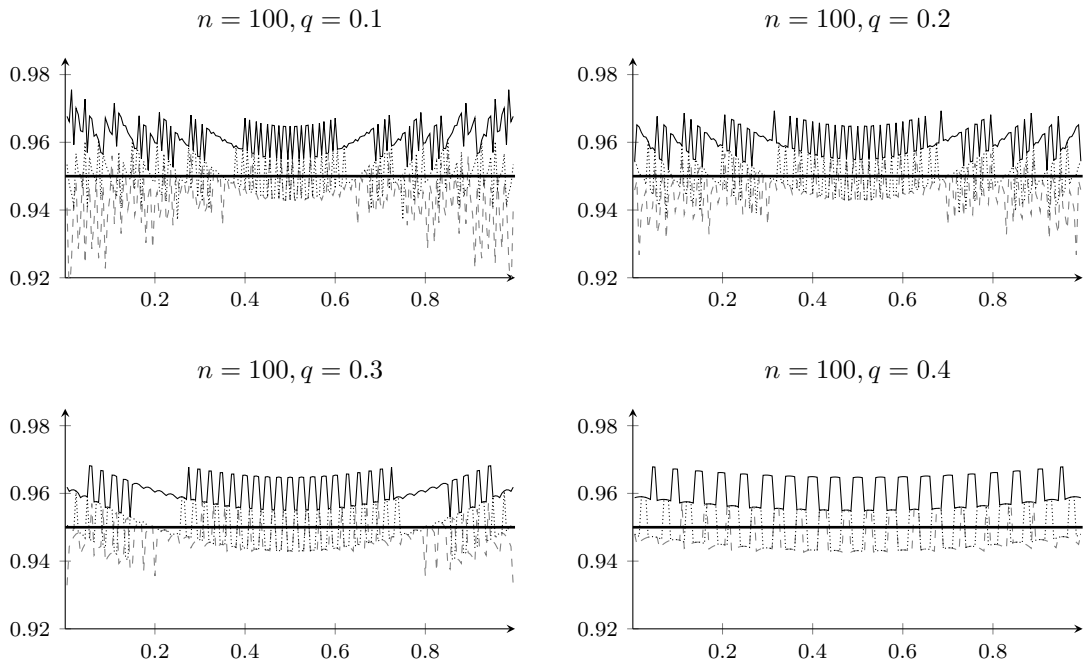

	\caption{Probability of coverage for $n=100$ and $\delta=0.95$}\label{fig:1}
	\begin{center}
		\begin{subfigure}[h]{7truecm}
			\centering
			\include{n100-q01-CP}
		\end{subfigure}
		\quad
		\begin{subfigure}[h]{7truecm}
			\centering
			\include{n100-q02-CP}
		\end{subfigure}
		\\
		\begin{subfigure}[h]{7truecm}
			\centering
			\include{n100-q03-CP}
		\end{subfigure}
		\quad
		\begin{subfigure}[h]{7truecm}
			\centering
			\include{n100-q04-CP}
		\end{subfigure}
	\end{center}
\end{figure}

\begin{figure}[H]
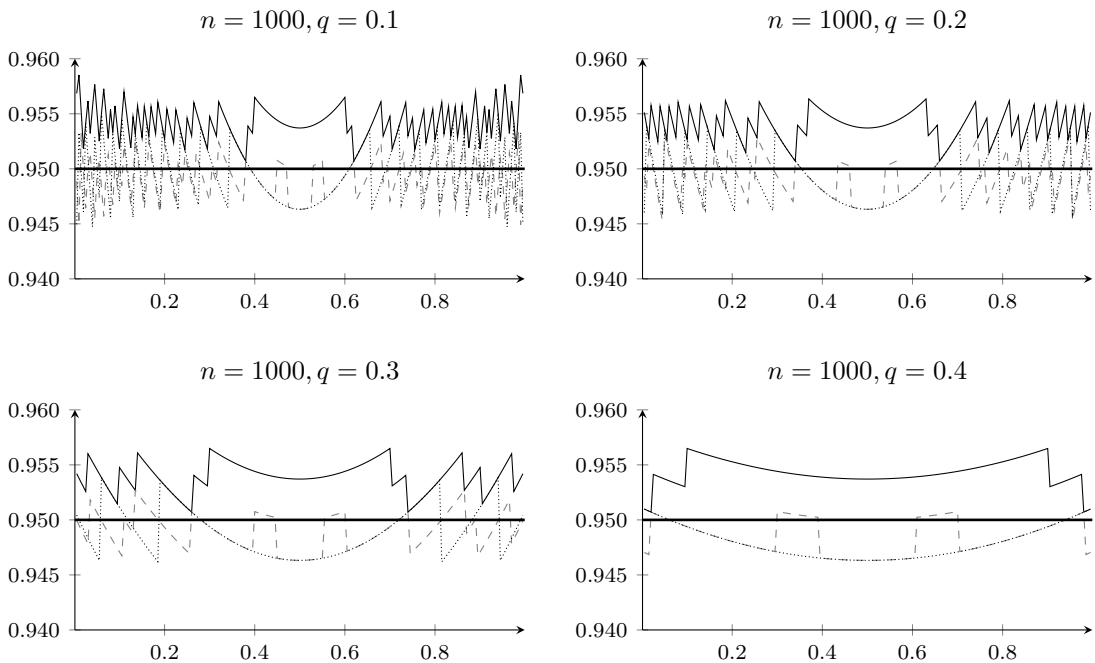

\caption{Probability of coverage for $n=1000$ and $\delta=0.95$}\label{fig:2}
	\begin{center}
		\begin{subfigure}[h]{7truecm}
			\centering
			\include{n1000-q01-CP}
		\end{subfigure}
		\quad
		\begin{subfigure}[h]{7truecm}
			\centering
			\include{n1000-q02-CP}
		\end{subfigure}
		\\
		\begin{subfigure}[h]{7truecm}
			\centering
			\include{n1000-q03-CP}
		\end{subfigure}
		\quad
		\begin{subfigure}[h]{7truecm}
			\centering
			\include{n1000-q04-CP}
		\end{subfigure}
	\end{center}
\end{figure}

The coverage probabilities of the $(WP)$ and $(AP)$ confidence intervals are shown in Figures \ref{fig:1} and \ref{fig:2} for $n=100$ and $n=1000$ (dotted and dashed lines, respectively). The confidence level is assumed to be $0.95$. These coverage probabilities are calculated, not simulated.


Proposed $(CP)$ c.i. keep the nominal confidence level and the risk of error in statement is not greater than $1-\gamma$. Unfortunately asymptotic ``confidence intervals'' do not satisfy the Neyman (1934) definition. The probability of coverage is less than the nominal confidence level, i.e. the risk of erroneous statement is greater than $1-\gamma$ and remains unknown. Hence in what follows we consider only $(CP)$ confidence intervals.

\section{The length of the exact confidence interval}

Let us consider  the length of the $CP$ confidence interval. For $x$ observed \Y answers to the questionnaire we have
$$l(x,q,n)=
\begin{cases}
\RightCi,&\hbox{if $\condL$},\\
\LeftCi,&\hbox{if $\condR$},\\
	\DeltaCi,&\hbox{otherwise.}\\
\end{cases}
$$
Recall that $u=n(x(2q-1)+(1-q))$.

The length of the c.i. is a random variable. It depends on $q$, $n$ and $x$. There are at least three approaches to the problem of minimizing the length of the c.i.:
\begin{penum}
	\item  minimizing for each $x$,
	\item  minimizing expected length,
	\item  almost sure  minimizing .
\end{penum}
Minimizing the length for observed $x$ relays on changing probabilities of over- and underestimation. This method was widely discussed by Zieli\`{n}ski (2010, 2017). To obtain the shortest c.i. in Neyman sense, i.e. controlling the probability of coverage, randomization is needed. In what follows we consider equitailed c.i. and confine ourselves to the problem of minimizing expected length and the almost sure minimization.

Note that the length  decreases as $n$ increases, hence the problem of minimizing the length is equivalent to finding appropriate $q$, i.e. the probability of a positive answer to neutral question. In what follows sample size $n$ is treated as a given number, and will be omitted in notations.

{\bf Minimizing expected length.} The problem to be solved may be written in the following way
	$$q^*_e=\underset{q\in Q}{\arg\min}\sup_{\pi\in\Pi}E_\pi l(Z,q),$$
	where $Q$ and $\Pi$ are acceptable sets for $q$ and $\pi$ accordingly. Without prior knowledge of $q$ and $\pi$ the set $Q=\langle 0,0.5)$ (under prior assumption that $q<0.5$) and $\Pi=(0,1)$.

{\bf Almost sure minimizing of length.} The problem to be solved may be written in the following way
	$$q^*_{\delta}=\underset{q\in Q}{\arg\max}\inf_{\pi\in\Pi}P_\pi\left\{l(Z,q)\leq\delta\right\},$$
	where $\delta$ is a given number chosen in advance: $\delta$ should be small.

\bigskip

It is easy to  see that for $Q=\langle 0,0.5)$ the minimal length with respect to $q$ is obtained for $q=0$, which is equivalent to nonasking the neutral question. Such a questionnaire (without a neutral question) is useless for our purposes. Hence we have to introduce a limitation for the probability $q$.

(Tan et al., 2009) introduced a notion of the degree of privacy protection  through probabilities $$P_{\pi,q}\left\{Y=1|Z=1\right\} \text{ and } P_{\pi,q}\left\{Y=1|Z=0\right\}.$$ These probabilities are connected with the safety of the interviewee of non-discovering her/his positive answer to the sensitive question. In CM model these probabilities are as follows (simple application of Bayes theorem)
$$\begin{aligned}
	p_{1,1}=P_{\pi,q}\left\{Y=1|Z=1\right\}&=\frac{\pi q}{\pi q+(1-\pi)(1-q)},\\
	p_{1,0}=P_{\pi,q}\left\{Y=1|Z=0\right\}&=\frac{\pi(1-q)}{\pi(1-q)+(1-\pi)q}.\\
\end{aligned}$$
These probabilities should be appropriately small, so that  they do not exceed a given value $\gamma \in (0,1)$. Note that for all $q\in(0,0.5)$ and for all $\pi\in (0,1 )$ holds $p_{1,0}\geq p_{1,1}$ (see Figure 3).
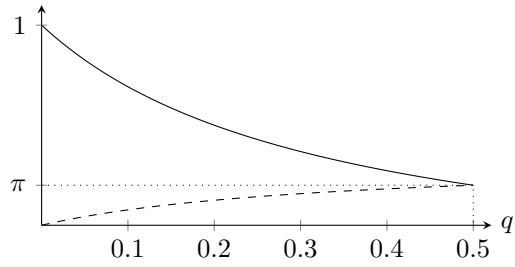
\begin{figure}[H]
\begin{center}
	\caption{Privacy protection versus $q$.}\label{fiq:3}
	\begin{tikzpicture}
	\def\pilevel{0.2}
	\def\gammalevel{0.6}
	\def\invgammalevel{(1-\gammalevel)*\pilevel/((1-\gammalevel)*\pilevel+(1-\pilevel)*\gammalevel)}
	\begin{axis}[
		axis x line=middle,
		axis y line=left,
		ymax=1.1, ymin=0,
		xmin=0, xmax=0.521,
		height=4.5cm,
		width=7.5cm,
		yticklabels={$\pi$,1},
		ytick={\pilevel,1},
		xlabel={$q$},
		 xlabel style={at={(ticklabel* cs:1)}, anchor=west },
		]
		\addplot 
		[domain=0:0.5,black,samples=250,dashed]{\pilevel*x/(\pilevel*(1-x)+(1-\pilevel)*x)};
		\addplot
		[domain=0:0.5,black,samples=250]{\pilevel*(1-x)/(\pilevel*(1-x)+(1-\pilevel)*x)};
		\addplot[dotted] coordinates {(0,\pilevel) (0.5,\pilevel) (0.5,0)};
	\end{axis}
\end{tikzpicture}
\end{center}
\end{figure}

Therefore we are interested in $q<0.5$  such that 
\begin{equation}\label{equ:1}
		p_{1,0}=P_{\pi,q}\left\{Y=1|Z=0\right\}\leq\gamma\quad\text{for }\pi\in\Pi.
\end{equation}

Simple algebra gives the following condition for $q$:
\begin{equation}\label{equ:2}q(\pi;\gamma)\leq q<0.5\quad\text{for }\pi\in\Pi,\end{equation}
where $q(\pi;\gamma)=\frac{\pi(1-\gamma)}{\gamma(1-2\pi)+\pi}$. Since $q(\gamma,\gamma)=0.5$ for all $\pi\in\Pi$, the above condition (\ref{equ:2}) holds for $\gamma>\pi$. It means that the maximal privacy protection (i.e. minimal $\gamma$ to be chosen) is limited  by the percentage of population who committed  socially stigmatizing crimes.  Hence the problem of minimizing the length, assuming $\pi\leq\pi_0$, for a given $\pi_0\in (0,1)$, is well defined for $q\in\langle q(\pi_0;\gamma), 0.5)$. The privacy protection criterion is satisfied for $\gamma>\pi_0$. In what follows we assume that $\Pi=(0,\pi_0\rangle$ and  $Q=\langle q(\pi_0;\gamma), 0.5)$.

Note that $\Pi$, as well as $Q$, does not depend on sample size $n$. So the length of the c.i. may be minimized by choosing appropriate sample size. 

Let $d\in(0,1)$ be a given number. We would like to find sample size which gives the c.i. of the length not greater than $d$. In particular, we are interested in c.i. covering estimated value of $\pi$. There are two approaches to the problem: find minimal $n$ such that

\begin{enum}
	\item $E^\c_{\pi}l(Z_n,q,n)\leq d$ for all $\pi\in\Pi$ and for all $q\in Q$, \par\vspace{6pt} 
	where $E^{\c}_{\pi}l(Z_n,q,n)=\sum_{x\in \c} l(x,q,n)P_\pi\{Z_n=x\}$ denotes the expected length of c.i. covering estimated value of $\pi$. The set   $\c=\{x\in\{0,1,\ldots,n\}:\ \pi_L(x;\delta)<\pi<\pi_R(x;\delta))\}$ includes those values of the variable $Z_n$ for which the c.i. covers $\pi$. \vspace{6pt} 
    \item  $P^\c_{\pi}\left\{l(Z_n,q,n)\leq d\right\}\geq 1-\lambda$ for a given probability $1-\lambda$ and for all $\pi\in\Pi$ and for all $q\in Q,$\par\vspace{6pt}   
    where   $\delta\cdot  P^\c_{\pi}\left\{l(Z_n,q,n)\leq d\right\}=\sum_{x\in \c} P_\pi\{Z_n=x\}\Jdn(l(x,q,n)\leq d)$
    denotes the probability that the length of c.i. covering estimated value of $\pi$ does not exceed $d$. The function $\Jdn(p)$ is equal to one if a logic value of $p$ is true and zero otherwise.
\end{enum}
In the  first approach we want the average length of the c.i. covering the estimated value of $\pi$ to be less than given $d$. In the second approach we want  the length of at least $(1-\lambda)\%$ of c.i. covering estimated $\pi$ to be less than given $d$. Let us note that we have at least $\delta\%$ of intervals covering the unknown parameter $\pi$ and for infinitely large sample size $n$ the defined value $P^\c_{\pi}\left\{l(Z_n,q,n)\leq d\right\}$ is equal to one.\par
\begin{figure}[H]
	\begin{center}
		\caption{Expected length versus $q$ with respect to $\pi$ under the condition that $\pi$ is in the confidence interval.}\label{graph:1} 
		\includegraphics[scale=0.8]{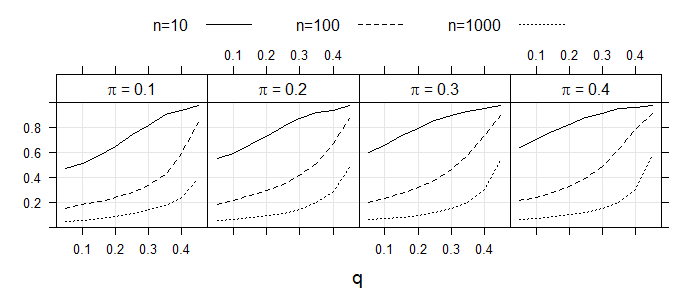}
	\end{center}
\end{figure}
\begin{figure}[H]
	\begin{center}
		\caption{Expected length versus $\pi$ with respect to $q$ under the condition that $\pi$ is in the confidence interval.}\label{graph:2}
		\includegraphics[scale=0.8]{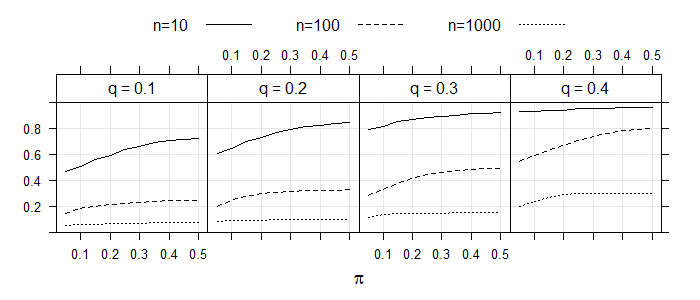}
	\end{center}
\end{figure}

Consider the first approach. The analysis of  $E^\c_{\pi}l(Z_n,q,n)$ shows that (see Figures \ref{graph:1}, \ref{graph:2})
\begin{penum}
	\item for each $q$ and $n$ it increases as $\pi$ increases for $\pi\in(0,0.5)$,
	\item for each $\pi$ and $n$ it increases as $q$ increases for $q\in(0,0.5)$.
\end{penum}
Hence it is enough to find sample size $n$ such that $E^\c_{\pi_0}l(Z_n,q(\pi_0,\gamma),n)\leq d.$
For a given $\pi_0$, $\gamma$ and $d$ the solution may be found numerically. 

In Table \ref{tab:1} some exemplary minimal sample sizes are given for confidence level $\delta=0.95$ and privacy protection $\gamma=0.5$.
\begin{table*}[h]
		\caption{The smallest $n$ that $E^\c_{\pi_0}l(Z_n,q(\pi_0,\gamma),n)\leq d$.}\label{tab:1}
		\centering
		\begin{tabular}{lrr}
			\hline
			\tb $\pi_0$ & $d=0.05$ & $d=0.06$ \\ 
			\hline
			 0.1 & 1326 & 929\\ 
			0.2 & 3428 & 2388 \\ 
			0.3 & 8557 & 5955 \\ 
			0.4 & 34862 &24206  \\ 
			\hline
			\multicolumn{3}{c}{\tb Note: $\pi_0=q(\pi_0,\gamma)$ for $\gamma=0.5$ } \\ 
			\hline
		\end{tabular}
		\end{table*}

Now consider the second approach, i.e. we want to find sample size $n$ such that
$$P^\c_{\pi}\left\{l(Z_n,q,n)\leq d\right\}\geq 1-\lambda$$
for a given $d$ and $\lambda$.
The analysis of $P^\c_{\pi}\left\{l(Z_n,q,n)\leq d\right\}$ shows that  (see Figures \ref{fig:4} and \ref{fig:5})
\begin{penum}
	\item for every $d$, $q$ and $n$ it decreases in $\pi$ (for $\pi\in(0,0.5)$),
	\item for every $d$, $\pi$ and $n$ it does not decrease in $q$ (for $q\in(0,0.5)$).
\end{penum}
\begin{figure}[H]
	\begin{center}
	\caption{$P_{\pi}^\c\{l(Z,q)\leq d\}$ versus $\pi$. The case for  $n=1000$ and $\delta=0.95$}\label{fig:4}
	\begin{subfigure}[h]{6truecm}
		\centering
		\begin{tikzpicture}
	\begin{axis}[
		myfigureformat-pivsq,
		width=5.5cm,
		title={$d=0.05$},
		xlabel=$\pi$,
		xmin=0,
		xmax=0.1,
		ymax=1.08,
		xtick={0,0.02,0.05,0.08},
		xticklabels={0,0.02,0.05,0.08},
		legend pos=outer north east, 
		legend style={draw=none},
		xlabel style={at={(ticklabel* cs:1)}, above=15pt,anchor=west },
		ylabel style={at={(ticklabel* cs:1)},anchor=south west,rotate=-90}
		]
		\addplot[color=black] coordinates {
(0.001,0.966)
(0.006,0.92)
(0.011,0.838)
(0.016,0.719)
(0.021,0.571)
(0.026,0.416)
(0.031,0.276)
(0.036,0.166)
(0.041,0.09)
(0.046,0.044)
(0.051,0.019)
(0.056,0.008)
(0.061,0.003)
(0.066,0.001)
(0.071,0)
(0.076,0)
		};
		\addplot[dashed,color=black] coordinates {
(0.001,0.922)
(0.006,0.852)
(0.011,0.752)
(0.016,0.626)
(0.021,0.488)
(0.026,0.352)
(0.031,0.235)
(0.036,0.145)
(0.041,0.082)
(0.046,0.043)
(0.051,0.02)
(0.056,0.009)
(0.061,0.004)
(0.066,0.001)
(0.071,0)
(0.076,0)
		};
		\addplot[densely dotted,color=black] coordinates {
(0.001,0.808)
(0.006,0.712)
(0.011,0.6)
(0.016,0.48)
(0.021,0.363)
(0.026,0.259)
(0.031,0.174)
(0.036,0.11)
(0.041,0.065)
(0.046,0.036)
(0.051,0.019)
(0.056,0.009)
(0.061,0.004)
(0.066,0.002)
(0.071,0.001)
(0.076,0)
		};
	\addplot[dash pattern=on 0.5pt off 1pt on 4pt off 1pt,color=black] coordinates {
(0.001,0.045)
(0.006,0.042)
(0.011,0.04)
(0.016,0.037)
(0.021,0.034)
(0.026,0.032)
(0.031,0.03)
(0.036,0.028)
(0.041,0.026)
(0.046,0.024)
(0.051,0.022)
(0.056,0.021)
(0.061,0.019)
(0.066,0.018)
(0.071,0.016)
(0.076,0.015)
	};
	\end{axis}
\end{tikzpicture}
	\end{subfigure}
	\quad
	\begin{subfigure}[h]{8truecm}
		\centering
		\begin{tikzpicture}
	\begin{axis}[
		myfigureformat-pivsq,
		width=7.5cm,
		title={$d=0.06$},
		xlabel=$\pi$,
		xmin=0,
		xmax=0.16,
		ymax=1.08,
		xtick={0,0.03,0.06,0.09,0.12,0.15},
		xticklabels={0,0.03,0.06,0.09,0.12,0.15},
		legend pos=outer north east, 
		legend style={draw=none},
		 xlabel style={at={(ticklabel* cs:1)}, above=15pt,anchor=west },
		ylabel style={at={(ticklabel* cs:1)},anchor=south west,rotate=-90}
		]
		\addplot[color=black] coordinates {
(0.001,1)
(0.006,1)
(0.011,1)
(0.016,1)
(0.021,1)
(0.026,1)
(0.031,1)
(0.036,1)
(0.041,1)
(0.046,1)
(0.051,0.999)
(0.056,0.995)
(0.061,0.988)
(0.066,0.97)
(0.071,0.937)
(0.076,0.88)
(0.081,0.797)
(0.086,0.687)
(0.091,0.56)
(0.096,0.428)
(0.101,0.305)
(0.106,0.201)
(0.111,0.123)
(0.116,0.07)
(0.121,0.037)
(0.126,0.018)
(0.131,0.008)
(0.136,0.003)
(0.141,0.001)
(0.146,0)
		};
 \addlegendentry{$q=0.10$};	
		\addplot[dashed,color=black] coordinates {
(0.001,0.999)
(0.006,0.997)
(0.011,0.99)
(0.016,0.975)
(0.021,0.945)
(0.026,0.894)
(0.031,0.816)
(0.036,0.711)
(0.041,0.586)
(0.046,0.453)
(0.051,0.327)
(0.056,0.22)
(0.061,0.137)
(0.066,0.079)
(0.071,0.042)
(0.076,0.021)
(0.081,0.01)
(0.086,0.004)
(0.091,0.002)
(0.096,0.001)
(0.101,0)
(0.106,0)
(0.111,0)
(0.116,0)
(0.121,0)
(0.126,0)
(0.131,0)
(0.136,0)
(0.141,0)
(0.146,0)
		};
	 \addlegendentry{$q=0.12$};
		\addplot[densely dotted,color=black] coordinates {
(0.001,0.93)
(0.006,0.877)
(0.011,0.803)
(0.016,0.708)
(0.021,0.598)
(0.026,0.48)
(0.031,0.365)
(0.036,0.263)
(0.041,0.178)
(0.046,0.113)
(0.051,0.068)
(0.056,0.038)
(0.061,0.02)
(0.066,0.01)
(0.071,0.005)
(0.076,0.002)
(0.081,0.001)
(0.086,0)
(0.091,0)
(0.096,0)
(0.101,0)
(0.106,0)
(0.111,0)
(0.116,0)
(0.121,0)
(0.126,0)
(0.131,0)
(0.136,0)
(0.141,0)
(0.146,0)
		};
	 \addlegendentry{$q=0.15$};
	\addplot[dash pattern=on 0.5pt off 1pt on 4pt off 1pt,color=black] coordinates {
(0.001,0.052)
(0.006,0.048)
(0.011,0.045)
(0.016,0.042)
(0.021,0.04)
(0.026,0.037)
(0.031,0.034)
(0.036,0.032)
(0.041,0.03)
(0.046,0.028)
(0.051,0.026)
(0.056,0.024)
(0.061,0.022)
(0.066,0.021)
(0.071,0.019)
(0.076,0.018)
(0.081,0.016)
(0.086,0.015)
(0.091,0.014)
(0.096,0.013)
(0.101,0.012)
(0.106,0.011)
(0.111,0.01)
(0.116,0.009)
(0.121,0.008)
(0.126,0.008)
(0.131,0.007)
(0.136,0.006)
(0.141,0.006)
(0.146,0.005)
	};
 \addlegendentry{$q=0.45$};
	\end{axis}
\end{tikzpicture}
	\end{subfigure}
	\end{center}
\end{figure}
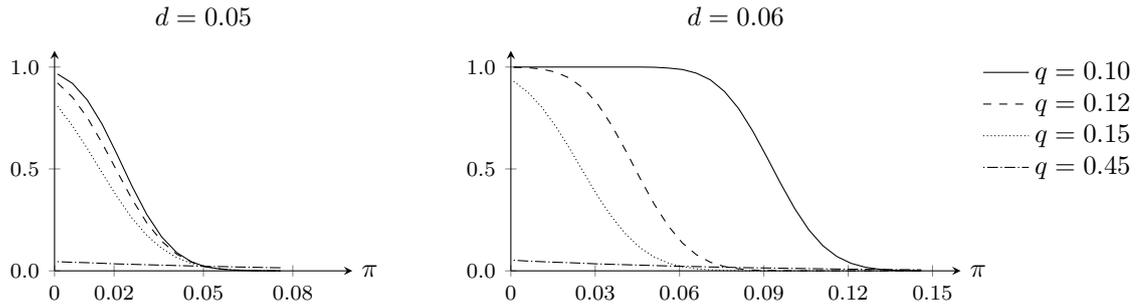
\begin{figure}[H]
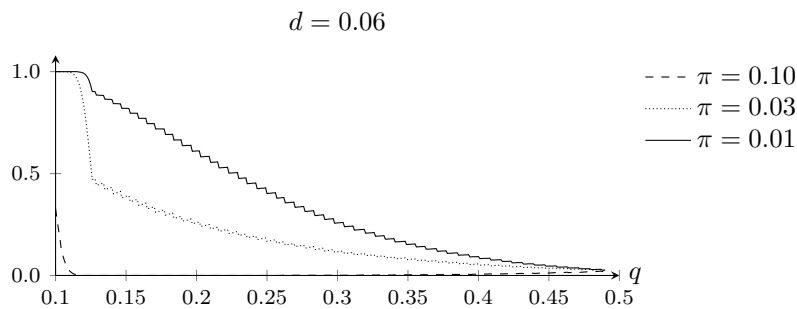

	\begin{center}
			\caption{$P_{\pi}^\c\{l(Z,q)\leq d\}$ versus $q$. The case for  $n=1000$ and $\delta=0.95$}\label{fig:5}
			\include{probd1000d06}
	\end{center}
\end{figure}
Monotonicity of  $P^\c_{\pi}\left\{l(Z_n,q,n)\leq d\right\}$ in $q$  is disturbed due to the discreteness of the observed variable $Z_n$.  However  the probability  is  generally decreasing in $q$  (see Figure \ref{fig:5}). This distinctive feature allows us to recommend that it is enough to find sample size $n$ such that $P^\c_{\pi_0}\left\{l(Z,q(\pi_0;\gamma))\leq d\right\}\geq 1-\lambda.$ For a given $\pi_0$, $\gamma$, $d$ and $\lambda$ the solution may be found numerically. In Table \ref{tab:2} some exemplary minimal sample sizes are given for confidence level $\delta=0.95$, privacy protection $\gamma=0.5$ and $\lambda=0.01$ and $\lambda=0.05$.
\begin{table*}[h]
	\caption{The smallest $n$ that $P^\c_{\pi_0}\left\{l(Z,q(\pi_0;\gamma))\leq d\right\}\geq 1-\lambda$}\label{tab:2}
	\centering
	\begin{tabular}{lrrrr}
		\hline
		\tb& \multicolumn{2}{c}{$d=0.05$} & \multicolumn{2}{c}{$d=0.06$} \\ 
		\cline{2-5}
		\tb$\pi_0$ &$\lambda=0.01$ & $\lambda=0.05$ & $\lambda=0.01$ & $\lambda=0.05$ \\ 
		\hline
		0.1 & 1570 & 1551 & 1111 & 1094 \\ 
		0.2 & 3861 & 3845 & 2699 & 2686 \\ 
		0.3 & 9508 & 9499 & 6623 & 6615 \\ 
		0.4 & 38576 & 38572 & 26819 & 26815 \\ 
		\hline
			\multicolumn{5}{l}{\tb Note: $\pi_0=q(\pi_0,\gamma)$ for $\gamma=0.5$ } \\ 
		\hline
	\end{tabular}
\end{table*}

As it may be expected the sample size increases when requirements for the length of c.i. increase, i.e. $d$ decreases. Also, to obtain c.i. of a given length for $\pi$ which is prior smaller, i.e. $\pi_0$ is smaller, the smaller sample size is needed. It is interesting, that a slight increase of sample size gives a higher percentage of c.i. of a given length covering unknown value of $\pi$  (probability $\lambda$ is smaller, i.e. $1-\lambda$ is greater).

\color{black}
\section{Conclusions}
In the paper the new confidence interval for the fraction of sensitive questions is proposed. Recall that c.i. were proposed by (Neyman, 1934). He defined a c.i. as a method of estimation such that ``... the probability of an error in a statement of this sort being equal to or less than $1-\varepsilon$, where $\varepsilon$ is any number $0<\varepsilon<1$, chosen in advance. The number $\varepsilon$ I call the confidence coefficient." (in our notation confidence level is $\gamma$). The new c.i. keeps the prescribed confidence level, while very popular asymptotic c.i. does not. 

A very important practical problem is a question about the sample size. We derived a minimal sample size fulfilling two criteria: average length and almost sure length. To derive these sample sizes we put restrictions on privacy protection, i.e. the probability of discovering the \Y answer to the sensitive question. This probability should be appropriately small, so the interviewee may  feel safe answering to the questionnaire. Also, we restrict ourselves to the rare phenomena, so we confine on sensitive questions with a small (given in advance) probability of a positive answer to the sensitive question.

We do not compare the length of our c.i. with asymptotic ones. The asymptotic c.i. must be shorter, since they do not keep prescribed confidence level: the real probability of coverage is less than the given confidence level. Hence the comparison of the lengths has no sense. Note that our confidence interval is very easy to calculate; even on each smartphone there is a spreadsheet application which calculates the quantiles of Beta distribution. Recall that asymptotic c.i., based on normal approximation, were useful in times when  computers were not easily available. So we recommend using our confidence interval in practice.

\includegraphics[scale=0.8]{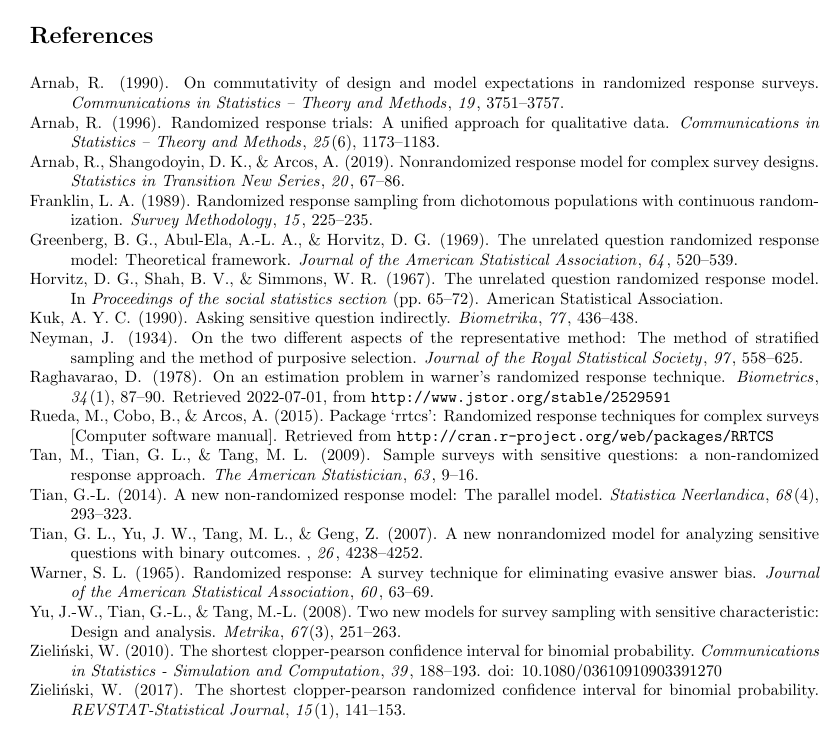}

\end{document}